# Robust High-Precision Time Transfer over 91-km Hollow-Core Fiber: Immunity to Dispersion and Nonlinearity

Bo Liu, Xinxing Guo, Jiang Chen, Huibo Hong, Qian Zhou, Xiang Zhang, Ru Yuan, Rongduo Lu, Tao Liu, Ruifang Dong, and Shougang Zhang

***Abstract*—To address the fundamental limitations imposed by chromatic dispersion and environmental susceptibility in Standard Single-Mode Fiber (SMF) for long-haul high-precision time transfer, we systematically explore the application potential of Hollow-Core Fiber (HCF) through comparative experiments. We designed a bidirectional time transfer platform enabling direct comparison between HCF and SMF links across distances of 91 km, 68 km, and 54 km. We quantitatively characterize the impact of critical non-reciprocal error sources—specifically the optical Kerr effect and chromatic dispersion—under varying laser power, wavelength drift, and environmental perturbations. Our results substantiate that HCF exhibits significantly suppressed dispersion (mean coefficient 3.4 ps/(nm·km)) and minimized environmental sensitivity compared to SMF. Notably, over the 91-km link, the HCF yields a signal-to-noise ratio (SNR) enhancement of >24 dB and confines time deviation to <80 ps—nearly an order-of-magnitude improvement over SMF (>600 ps)—while remaining virtually immune to power and wavelength fluctuations. Under 24-hour diurnal monitoring, the 68-km HCF link demonstrates exceptional robustness, with environment-induced time delay fluctuations (776 ps) reduced to just 24.5% of those in SMF (3166 ps). Consequently, the time transfer stability (TDEV) achieves 0.2 ps at an integration time of 1000 s, doubling the performance of SMF. These findings validate HCF as a superior transmission medium characterized by low latency, low nonlinearity, and thermal stability, paving the way for next-generation ultra-stable, long-haul time-frequency distribution networks.**

## I. INTRODUCTION

EMERGING applications in coherent radar arrays[1], deep-space navigation[2], and quantum networking[3] are pushing the requirements for time and frequency dissemination from the nanosecond regime toward the picosecond frontier. In this context, optical fiber links have established themselves as the preeminent medium for dissemination, exhibiting superior stability and noise suppression compared to satellite-based counterparts[4]. Fiber-optic time transfer based on Standard Single-Mode Fiber (SMF) has reached maturity, enabling picosecond-level stability over distances exceeding 100 km via optical loop-back and bidirectional comparison configurations [5, 6, 7, 8]. However, as demands for extended range and ultra-high precision intensify, the intrinsic physical limitations of SMF constitute a critical bottleneck[9, 10, 11].

In Wavelength Division Multiplexing (WDM)-based bidirectional schemes, precise synchronization relies on the assumption of strict path reciprocity to eliminate link latency[12, 13]. Yet, many practical systems necessitate distinct wavelengths for upstream and downstream signals. Owing to SMF's inherent chromatic dispersion (manifested as wavelength-dependent group velocity), these signals undergo unequal propagation delays; thus, residual non-reciprocal asymmetry remains even after common-mode perturbations are nominally cancelled via bidirectional measurement. This error, arising from the coupling between dispersion and laser wavelength drift, fundamentally constrains long-term stability and accuracy[14]. Further, SMF remains highly susceptible to environmental fluctuations; the interplay between the thermo-optic effect and thermal expansion in the silica core induces significant drift in both refractive index and optical path length, proving detrimental to high-precision time transfer applications[15, 16, 17].To circumvent these fundamental limitations, the rapid development of next-generation HCF technology presents a transformative opportunity for high-precision time-frequency transfer.

Firstly, concerning nonlinearity, the optical field in HCF is predominantly confined within the air core, drastically mitigating the Kerr nonlinearity associated with silica glass. Consequently, nonlinear phase noise and power-dependent delay drift are suppressed by several orders of magnitude, a characteristic indispensable for the long-haul transmission of high-power, broadband time-frequency signals.

Tailored microstructure designs enable HCFs to achieve flattened, near-zero dispersion profiles over broad spectral ranges. Crucially, the thermal sensitivity of dispersion in HCF is substantially lower than that of standard SMF (typical dispersion 17 ps/(nm·km) in C-band). This thermal stability significantly attenuates the impact of dispersion-wavelength drift coupling on long-term time transfer stability and accuracy[18].

Regarding optical attenuation, while standard SMF-28 typically exhibits losses of 0.2–0.25 dB/km at 1550 nm, state-of-the-art HCFs—specifically Nested Antiresonant Nodeless Fibers (NANF)—have broken this barrier[19]. Recent Double-Nested ANF (DNANF) designs have demonstrated a record-breaking loss of 0.09 dB/km at 1550 nm, surpassing the fundamental limit of solid-core silica fibers[20].

Finally, regarding environmental susceptibility, since light propagation is largely decoupled from the silica cladding, the influence of thermal expansion and the thermo-optic effect is minimized. This reduces the sensitivity of transmission delay to temperature fluctuations by one to three orders of magnitude compared to SMF[16, 18, 21]. Utilizing HCFs fabricated from ultra-low thermal expansion glass can further suppress the thermally induced delay coefficient by an additional three orders of magnitude[22]. Indeed, experimental demonstrations over kilometer-scale links have validated that HCF outperforms traditional SMF by more than an order of magnitude in terms of phase noise and frequency stability[23, 24].

While Hollow-Core Fiber (HCF) is heralded as a transformative candidate for overcoming the bottlenecks of current SMF-based links, rigorous comparisons between HCF and SMF within a unified high-precision time transfer framework remain scarce. Crucially, a quantitative understanding of how key physical parameters—specifically laser power, wavelength drift, chromatic dispersion, and thermal susceptibility—differentially impact the transfer performance of these two media is lacking. Consequently, direct experimental evidence delineating the comprehensive advantages and potential limitations of HCF under realistic operating conditions is still absent.

Here, we report a systematic study utilizing a reconfigurable bidirectional time transfer platform that enables direct, side-by-side characterization of HCF and SMF links under identical architectures. We quantitatively evaluate the impact of the aforementioned physical perturbations on time delay stability, received Signal-to-Noise Ratio (SNR), and transfer accuracy, thereby empirically validating the superior robustness of HCF in long-haul applications. Furthermore, by elucidating the distinct error mechanisms in both fibers, we analyze the necessity of dispersion compensation and active wavelength stabilization in traditional schemes. Finally, we provide critical design guidelines and parameter optimization strategies essential for the engineering implementation of next-generation, ultra-stable time-frequency distribution networks.

## II. THEORETICAL MODELING OF NON-RECIPROCAL ERRORS AND SYSTEM PERFORMANCE

### A.Modeling of Dispersion-Induced Wavelength-Delay Coupling

In a bidirectional time transfer scheme, denoting the upstream (A→B) and downstream (B→A) wavelengths as $\lambda_1$ and $\lambda_2$, the respective one-way propagation delays are given by:

$$\tau_{AB}=\tau_0+DL(\lambda_1-\lambda_0) \quad (1)$$

$$\tau_{BA}=\tau_0+DL(\lambda_2-\lambda_0) \quad (2)$$

Consequently, the differential bidirectional delay exhibits a residual non-reciprocal component directly proportional to the wavelength separation $\Delta\lambda=\lambda_1-\lambda_2$:

$$\Delta\tau_{nonrec,disp}=\tau_{AB}-\tau_{BA}=D\,L\,\Delta\lambda \quad (3)$$

In this work, "time deviation" (delay deviation) refers to the measured delay offset of the recovered timing observable; at first order, it is taken as the experimental manifestation of the residual wavelength-dependent propagation delay captured by $\Delta\tau_{nonrec,disp}$ .This relation provides a direct experimental mapping: the delay sensitivity to operating wavelength is governed by the cumulative dispersion $DL$, i.e., the slope $d(\Delta\tau)/d\lambda$ is proportional to DL[25, 26]. Therefore, the slope of the measured "time deviation versus wavelength" curve serves as a quantitative proxy for cumulative dispersion, and a near-zero slope indicates vanishing wavelength-to-delay coupling and strong immunity to wavelength fluctuations.

Using representative C-band parameters, standard SMF exhibits D_SMF on the order of $10^1$ ps/(nm·km), whereas HCF typically shows D_HCF on the order of a few ps/(nm·km)[21]. Under identical $L$ and $\Delta\lambda$, the dispersion-induced timing sensitivity ratio is predicted as

$$\frac{\Delta\tau_{SMF}}{\Delta\tau_{HCF}}=\frac{D_{SMF}}{D_{HCF}}\approx 3\text{-}5 \quad (4)$$

For a representative small detuning $\Delta\lambda \sim 10^{-1}$ nm and a long-haul distance L ~ $10^2$ km, the dispersion-induced delay excursion is expected to be on the order of $10^2$ ps in SMF, but only on the order of a few×$10^1$ ps in HCF. These quantitative estimates predict that (i) the delay–wavelength slope scales linearly with distance L, and (ii) for identical tuning conditions the slope and resulting delay excursion in SMF should exceed that in HCF by a few-fold factor (typically ~3–5).

Accordingly, a wavelength-sweep experiment over a controlled tuning window provides a direct and self-consistent test of the dispersion model through the measured slope $d(\Delta\tau)/d\lambda$ and its scaling with L.

### B. Comparative Analysis of Power-Dependent SNR in SMF and HCF Links

The optical Kerr effect in SMF renders the refractive index intensity-dependent, $n=n_0+n_2I$. With $I\propto P/A_{eff}$,optical-power variations perturb the effective refractive index and induce nonlinear phase accumulation[27]. Introducing the nonlinear coefficient $\gamma=\frac{2\pi n_2}{\lambda A_{eff}}$, the nonlinear phase shift is $\varphi_{NL}=\gamma P L_{eff}$. Consequently, power fluctuations $\Delta P$(in linear units) generate nonlinear phase noise $\Delta\varphi_{NL}\approx\gamma\Delta P L_{eff}$, which can be converted into timing jitter through system-dependent phase-to-time conversion in the phase-tracking and timing-extraction process; hence the Kerr-mediated delay perturbation can be written in the proportional form $\Delta\tau_{Kerr}\approx C_\tau\gamma L_{eff}\Delta P$, where, $C_\tau$ is determined by the receiver and timing-estimation algorithm.

$$\Delta\tau_{Kerr}\propto\gamma L_{eff}\Delta P \quad (5)$$

Importantly, the experimentally observed power-dependent delay deviation is not solely governed by fiber nonlinearity; it also inherits an intrinsic contribution from the SFP laser source. In practical SFP/DFB modules, tuning the launched optical power (typically via bias/current) unavoidably induces a concomitant shift of the emission wavelength through thermal drift and chirp. Linearizing around the operating point gives $\Delta\lambda_{SFP}\approx(d\lambda/dp_{dB})\Delta P_{dB}$ ,where $P_{dB}$ denotes optical power in dB units. In a bidirectional link, such source-induced wavelength drift contributes to the residual non-reciprocity only through a differential change of the wavelength separation, $\delta(\Delta\lambda) = \delta(\lambda_1 - \lambda_2)$ ,e.g., when the counter-propagating carriers are generated by non-identical SFP modules or experience unequal power tuning/thermal states[28].

Through the dispersion-induced wavelength–delay coupling derived in Section A, this produces an additional dispersion-mediated delay contribution

$$\Delta\tau_{\mathrm{SFP}\to\mathrm{disp}} \approx \mathrm{DL}\delta(\Delta\lambda) < \mathrm{DL}\left|\frac{\mathrm{d}\lambda}{\mathrm{dP_{dB}}}\right| |\Delta\mathrm{P_{dB}}| \qquad (6)$$

Here the upper bound corresponds to a conservative (worst-case) scenario in which the differential wavelength separation change is bounded by the magnitude of the source-induced drift.

Accordingly, the overall power sensitivity of the link delay can be interpreted as the superposition of a Kerr-mediated channel and a dispersion-mediated channel driven by the SFP source. For practical SFP modules, the source power-to-wavelength coupling typically lies in the "tens of pm per dB" regime, i.e., |dλ/dP_dB| ~ O (10) pm/dB, and can be experimentally upper-bounded by combining the independently measurable wavelength sensitivity $d(\Delta\tau)/d\lambda$ (from wavelength sweeping) and power sensitivity $d(\Delta\tau)/dP_dB$ (from power stepping).

With long-haul cumulative dispersion DL and moderate power steps |ΔP_dB| of order unity to a few dB, the dispersion-mediated term alone can yield delay excursions on the order of $10^2$ ps in SMF, whereas the corresponding excursions in HCF are expected to be substantially smaller due to its much lower *D*.

In parallel, nonlinear phase noise degrades the effective SNR. A compact expression highlighting the nonlinear penalty is

$$\mathrm{SNR}=\frac{(\mathrm{RP_{rec}})^2}{\sigma^2_{\mathrm{n,lin}}+\mathrm{K}(\gamma \mathrm{L_{eff}}\Delta \mathrm{P}\,)^2} \qquad (7)$$

where $\sigma^2_{n,lin}$ denotes the linear noise floor and *K* encapsulates system-dependent conversion factors. In this expression, $\Delta P$ denotes the linear power fluctuation (in W) relevant to Kerr-induced phase noise. This form separates two limiting regimes: at sufficiently high received power the nonlinear phase-noise term can become non-negligible and reduce the effective SNR margin, whereas at low received power the SNR is primarily limited by the linear noise floor and reduced tracking margin.

In long-haul operation, reduced SNR margin and power-induced delay perturbations are expected to jointly increase the likelihood of cycle slips and loss-of-lock, and this can be assessed via controlled power-variation measurements of SNR and delay deviation.

At a representative C-band operating point over a ~$10^2$ km link, HCF is expected to sustain a robust SNR margin due to its ultralow nonlinearity and low dispersion. In contrast, SMF is expected to exhibit reduced robustness against power-induced phase noise and power-to-delay coupling, particularly over long interaction lengths and under limited power margin.

Hollow-core fiber (HCF) fundamentally suppresses these power-induced impairments. First, guidance predominantly in air yields an ultralow nonlinear refractive index ($n_2^{air} \approx 3 \times 10^{-23} \mathrm{m}^2/\mathrm{W}$ )[29]compared with silica ( $n_2^{\mathrm{SMF}} \approx 2.5 \times 10^{-20} \mathrm{m}^2/\mathrm{W}$)[30], together with reduced optical-field overlap with solid material, resulting in a substantially smaller effective γ and hence strongly reduced Kerr-induced phase noise. Second, HCF also features a much smaller chromatic dispersion *D*, which suppresses the dispersion-mediated conversion of SFP-induced wavelength drift into timing errors. More generally, the effective nonlinearity can be expressed as $\gamma_{\mathrm{eff}}=\frac{2\pi \mathrm{n_{2,eff}}}{\lambda \mathrm{A_{eff}}}$, where $\mathrm{n_{2,eff}}$ accounts for modal overlap with nonlinear material; introducing a material-overlap factor $\mathrm{n_{mat}}$ $(0 \leq \mathrm{n_{mat}} \leq 1)$, one may write $\mathrm{n_{2,eff}} \approx \mathrm{n_{mat}} \mathrm{n_2^{silica}} + 1 - \mathrm{n_{mat}} \mathrm{n_2^{air}}$, such that $\mathrm{n_{mat}} \approx 1$ for SMF and $\mathrm{n_{mat}} \ll 1$ for HCF.

Therefore, HCF is expected to exhibit at least an order-of-magnitude lower effective nonlinear sensitivity than SMF, providing a physics-based basis for superior robustness against Kerr-induced phase noise and dispersion-mediated power-to-delay coupling in long-haul time-transfer links.

## III. EXPERMENTAL SYSTEM CONFIGURATION

The experimental setup consists of three primary segments: a local station, a remote station, and a reconfigurable fiber link. At the local station, an atomic clock provides a master 1 PPS/10 MHz reference to the Time Transfer Unit (TTU). This reference is encoded and modulates a tunable laser, whose output passes through a Variable Optical Attenuator (VOA)—for precise power regulation—and is launched into the fiber link via port 1→2 of an optical circulator. The return signal from the remote station is directed via port 2→3 to a photodetector (PD) for optoelectronic conversion. A decoding module then recovers the remote timestamp, which is compared against the local reference by an internal Time Interval Counter (TIC) to derive the delay measurement.

The remote station employs a symmetric architecture. The received optical signal is detected and decoded to recover the local timing information, which concurrently drives the remote TIC and disciplines a local oscillator to regenerate a synchronized 1 PPS/10 MHz signal. This regenerated clock, after phase adjustment via a Delay Compensator, modulates the remote laser for the return transmission. The platform enables seamless switching between HCF and SMF spools (nominal lengths: 54, 68, and 91 km) via optical patch cords. Note that while all spools are from a single manufacturer, minor parametric variations exist due to distinct production batches; detailed characterizations are provided in subsequent sections.

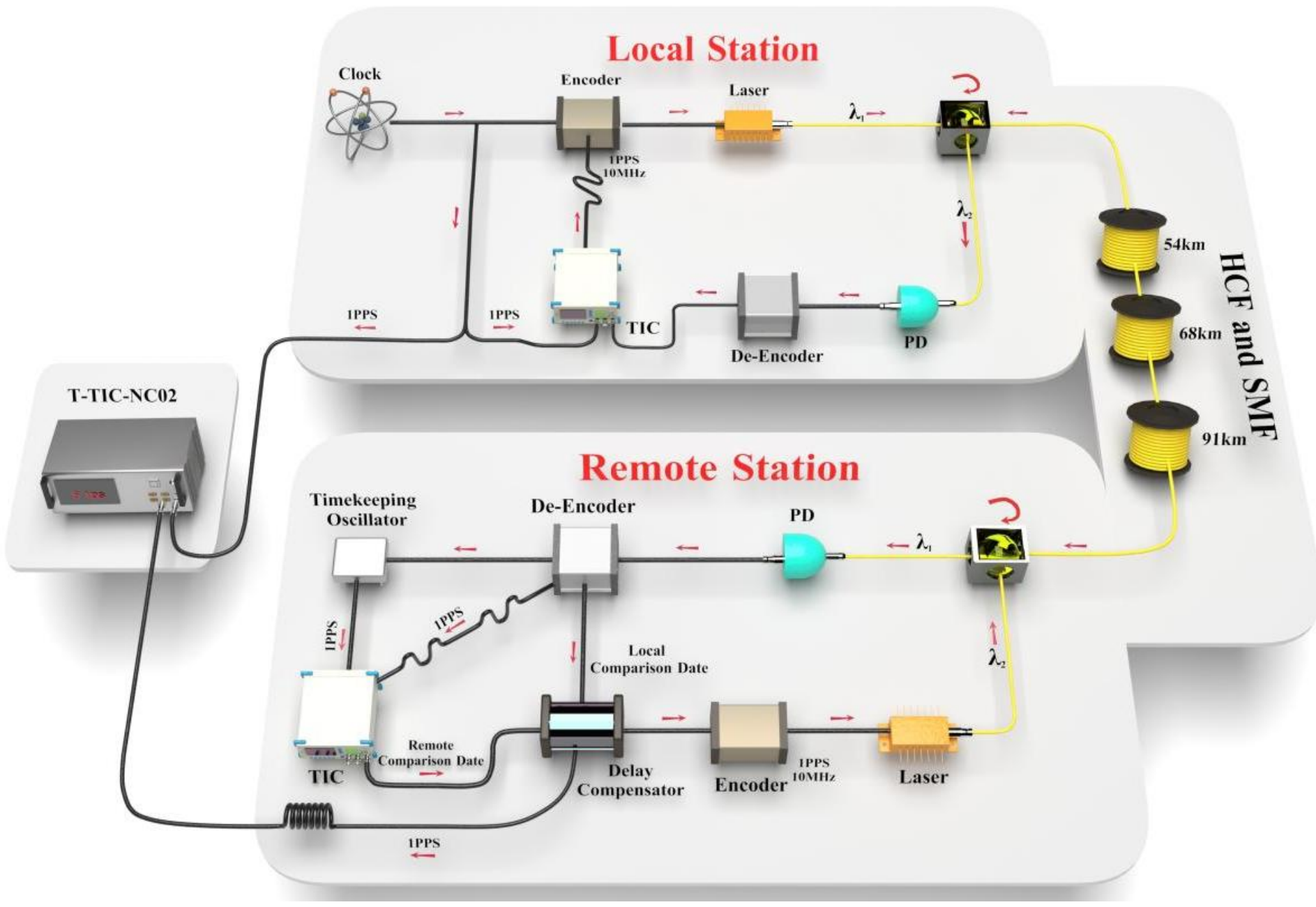

**Fig. 1.** Schematic illustration of the reconfigurable bidirectional time transfer experimental platform.

The system operation is orchestrated by a host computer via a USB interface using standard SCPI commands for synchronized control and data acquisition. Two automated protocols were implemented:

(1) Wavelength Sensitivity Test: The laser wavelength is linearly swept over a 120 pm range (rate: 20 pm/30 min) while clamping the output power at 0 dBm, with closed-loop feedback from a high-precision wavemeter.

(2) Power Sensitivity Test: The laser power is stepped through a pre-programmed sequence (e.g., 0, -5, -10 dBm), with data acquisition triggered upon power stabilization at each setpoint.

Crucially, all telemetry—including TIC readings, spectral data, and power levels—is time-stamped against a unified master clock and consolidated into a centralized dataset. This ensures rigorous temporal correlation between the control parameters and performance metrics, guaranteeing the fidelity of the comparative analysis.

## IV. EXPERIMENTAL RESULTS AND DISCUSSION

Prior to comparative time transfer experiments, we characterized the fundamental optical properties of both HCF and SMF links to establish a quantitative physical baseline for the subsequent performance analysis.

The comprehensive measurement results, presented in Fig. 2, reveal that HCF consistently surpasses conventional SMF across all critical transmission metrics, with the performance disparity amplifying systematically over distance. As illustrated in Fig. 2(a) and (b), the HCF maintains a remarkably stable attenuation coefficient of 0.17 dB/km—significantly lower than the 0.26–0.32 dB/km measured for the SMF spools. Consequently, over the 91-km link, the HCF incurs a total loss of only 15.81 dB compared to 28.58 dB for SMF, yielding a power budget margin advantage of nearly 13 dB. This substantial reduction in attenuation implies that for a fixed launch power, HCF can support substantially extended transmission reaches or accommodate higher splitting ratios in network architectures.

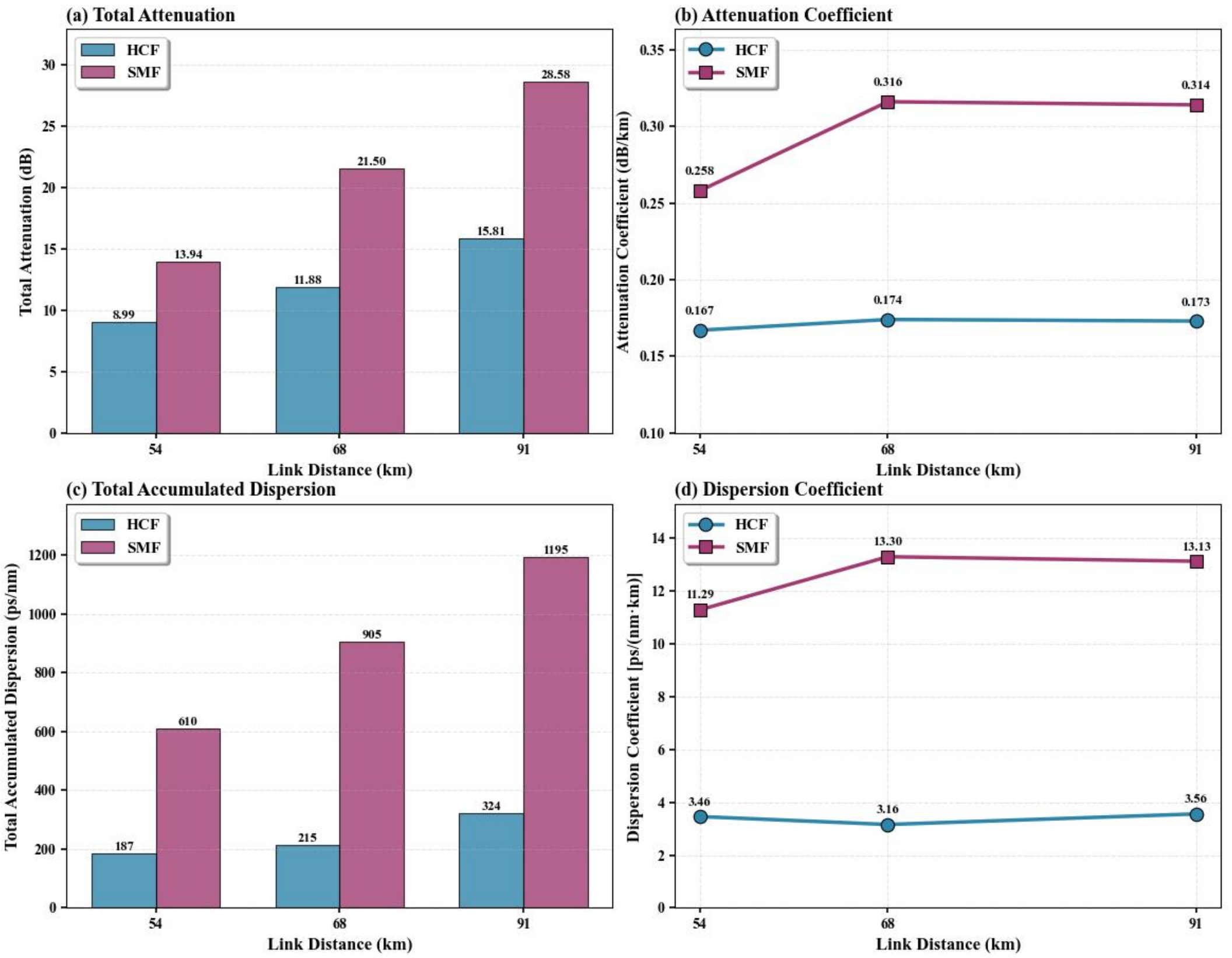

**Fig. 2.** Comparative transmission characteristics of HCF and SMF links at three distances (54 km, 68 km, and 91 km). (a) Total attenuation comparison showing HCF achieves 13 dB lower loss than SMF at 91 km. (b) Attenuation coefficient stability, with HCF maintaining ~ 0.17 dB/km versus SMF's 0.26–0.32 dB/km. (c) Total accumulated chromatic dispersion, demonstrating HCF's 3.7-fold reduction (324 ps/nm vs. 1195 ps/nm at 91 km). (d) Dispersion coefficient comparison, highlighting HCF's consistently low values of 3.1–3.6 ps/(nm·km) compared to SMF's 11–13 ps/(nm·km). The superior performance of HCF in both attenuation and dispersion domains establishes its fundamental advantage for ultra-stable time-frequency transfer.

The contrast is even more pronounced in the chromatic dispersion domain, as depicted in Fig. 2(c) and (d). The HCF exhibits a remarkably low dispersion coefficient ranging from 3.1 to 3.6 ps/(nm·km), whereas the SMF values span 11 to 13 ps/(nm·km)—representing a 3–4-fold reduction. At the 91-km distance, the total accumulated dispersion for HCF is merely 324 ps/nm, compared to 1195 ps/nm for SMF. These superior physical attributes fundamentally underpin the exceptional stability and accuracy of HCF in high-precision time transfer applications, as chromatic dispersion directly couples with laser wavelength drift to induce timing errors.

In operational environments, inevitable laser wavelength drift couples with the substantial chromatic dispersion of SMF, translating directly into severe timing errors. HCF, by virtue of its inherently low dispersion, offers a promising solution to mitigate this coupling. To validate this, we emulated spectral drift by actively tuning the laser wavelength over a 120 pm range; the experimental outcomes are summarized in Figs. 3 and 4.

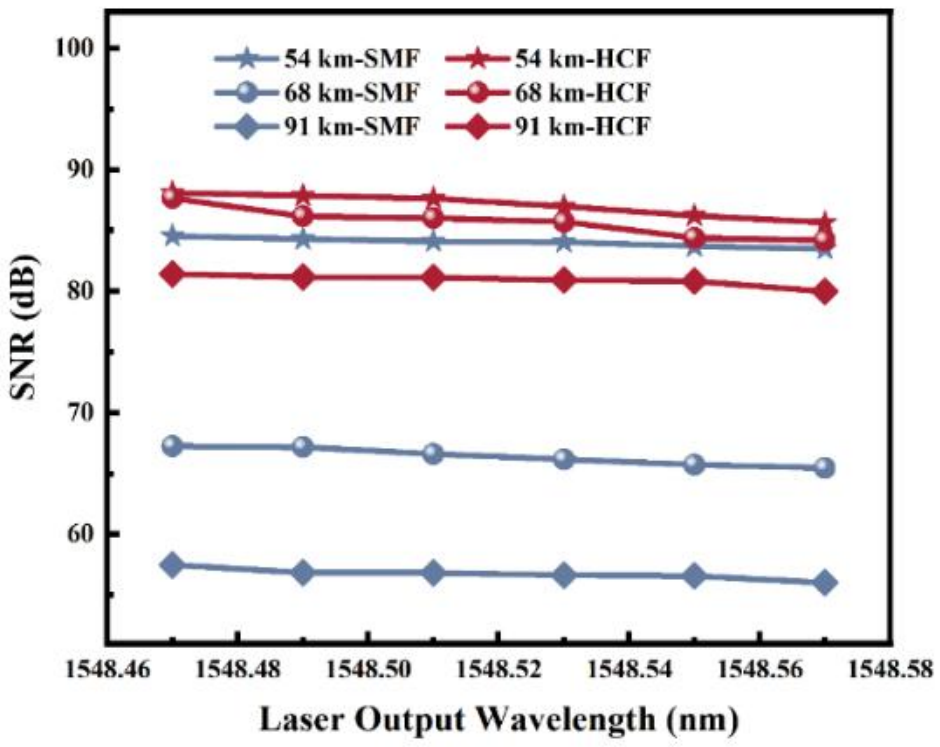

**Fig. 3.** Comparative SNR performance under wavelength tuning.

Fig. 3 compares the SNR evolution of HCF and SMF links at 54, 68, and 91 km versus laser output wavelength (1548.46-1548.58 nm). A slight monotonic decline in SNR is observed for both fiber types as the wavelength increases; this is attributed to the inherent output power roll-off of the tunable laser source rather than fiber dispersion or loss variations over this narrow spectral range (0.12 nm). Despite this source-induced fluctuation, the HCF link demonstrates superior robustness, maintaining an SNR > 80 dB. In stark contrast, the SMF link suffers severe degradation, dropping to ~ 57 dB at 91 km. This 20-dB performance gap is fundamentally governed by the nonlinear phase shift ($\emptyset_{NL} = \gamma P L_{eff}$). The air-guiding mechanism of HCF yields a negligible nonlinearity ($n_2 \approx 2.5 \times 10^{-20} m^2/W$), allowing the system to operate in a linear, noise-limited regime where SNR is maximized. Conversely, the SMF performance is capped by Kerr-induced nonlinear saturation (e.g., SBS or SPM), preventing SNR improvement even at shorter distances.

Fig. 4 provides experimental validation of the theoretical model discussed in Section A, quantifying the superior dispersion management of HCF. As derived in the theoretical analysis, the differential propagation delay is linearly proportional to the product of the chromatic dispersion coefficient ($D$) and the wavelength detuning ($\Delta\lambda$); thus, the gradient of the curves in Fig. 4 serves as a direct proxy for cumulative dispersion. Consistent with the theoretical projection that the dispersion coefficient of SMF (D_SMF) is inherently 3 to 5 times larger than that of HCF, the experimental data reveal a proportional disparity in delay sensitivity. Specifically, for the 91 km link, the SMF exhibits a steep gradient where a mere 120 pm wavelength shift induces a delay fluctuation exceeding 120 ps, with the total deviation escalating to nearly 700 ps. In stark contrast, the HCF maintains a near-zero slope—approximately 4 times lower than that of the SMF—resulting in a time deviation stably confined within 150 ps. These results definitively demonstrate that the dispersion-induced timing error in SMF is three- to fourfold that of HCF, confirming that the hollow-core medium's near-zero dispersion physics effectively suppresses systematic errors arising from wavelength instability, providing a critical safeguard for high-precision time transfer.

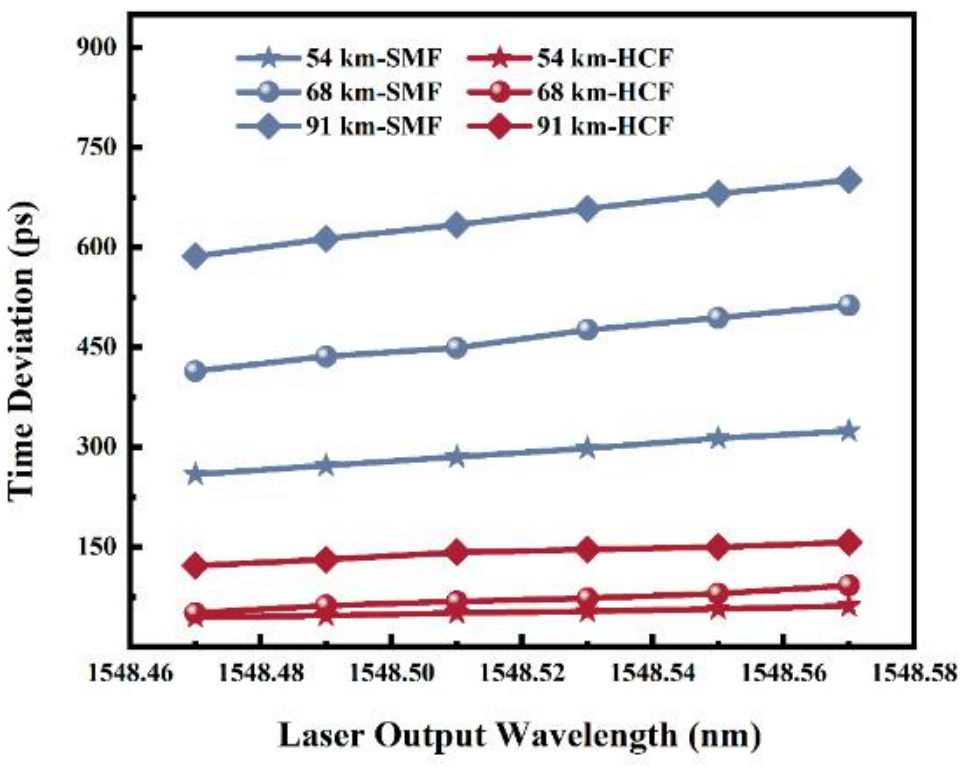

**Fig. 4.** Wavelength dependence of time deviation and dispersion characteristics.

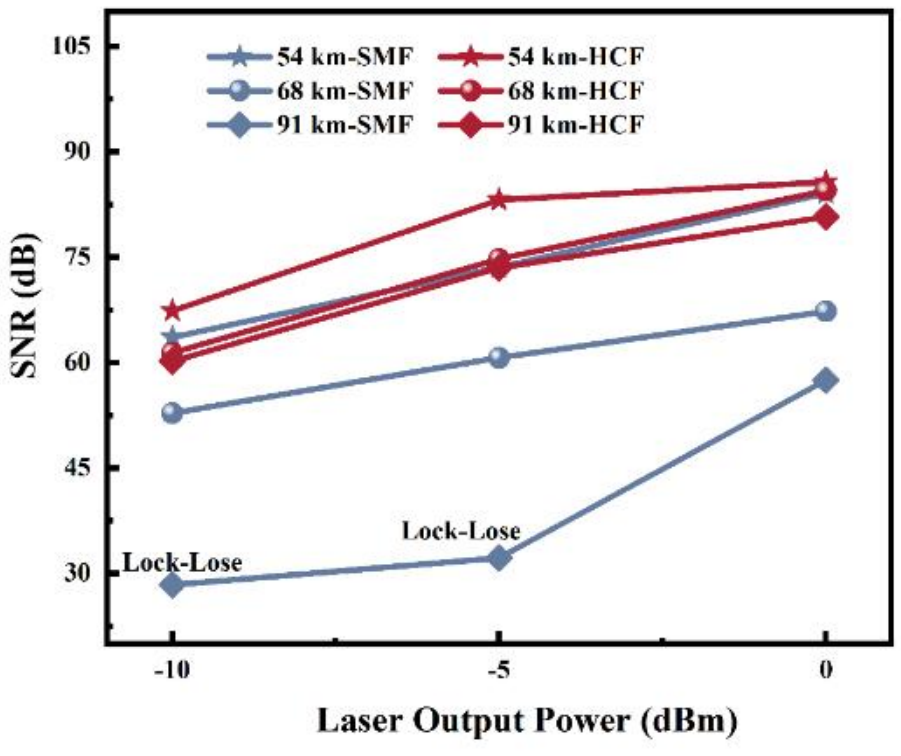

**Fig. 5.** Comparative SNR performance as a function of laser output power.

The optical Kerr effect constitutes a critical physical limitation in SMF links, where the intensity-dependent refractive index ($n = n_0 + n_2 I$) creates a deleterious coupling between optical power and propagation delay. Quantitative analysis reveals the sheer magnitude of this disparity: the nonlinear refractive index of the silica core is inherently three orders of magnitude higher than that of the hollow-core air medium. This massive theoretical gap manifests drastically in the experimental results shown in Figs. 5 and 6. In the 91-km SMF link, the accumulated nonlinear phase noise causes the SNR to plummet to ~ 28 dB at -10 dBm input optical power, cause the system to unlock. In stark contrast, the air-guiding mechanism of HCF suppresses this nonlinear penalty, maintaining a pristine SNR above 60 dB under identical conditions, demonstrating a superior > 30 dB margin. Consequently, while the time deviation in SMF escalates sharply to 1400 ps due to nonlinear phase noise, the HCF exhibits significant immunity to power variations, with deviation stably confined below 150 ps (an improvement of nearly an order of magnitude). These results definitively validate that HCF's air-guiding mechanism fundamentally circumvents the nonlinear power-delay coupling that severely degrades long-distance SMF links.

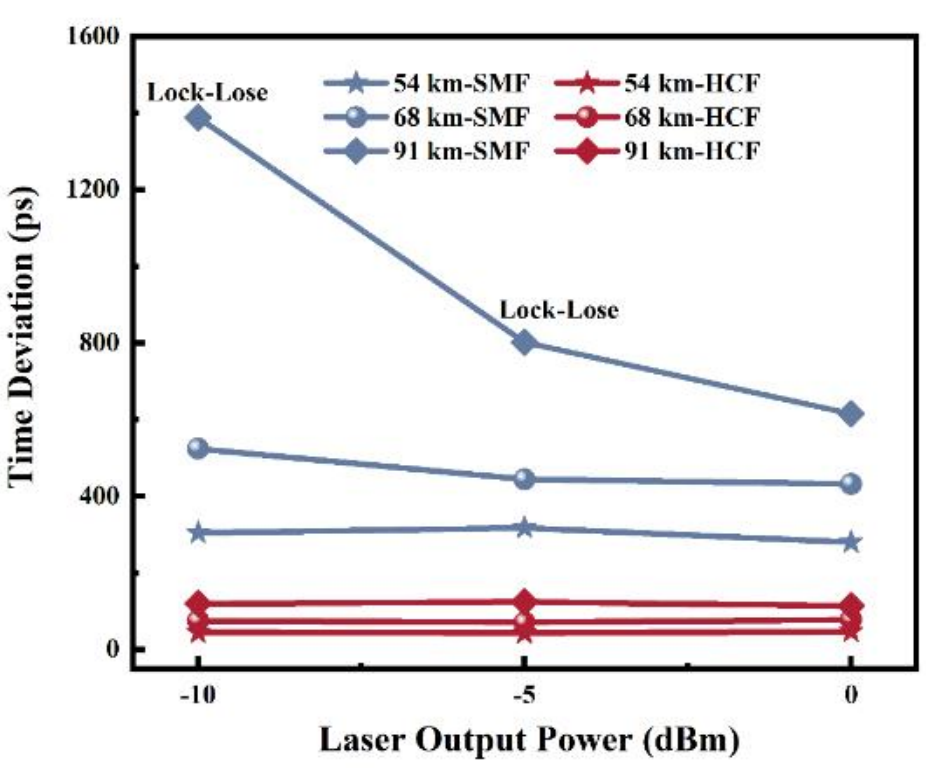

**Fig. 6.** Fiber-length-dependent time deviation of HCF and SMF at three laser output power levels.

Figs. 7(a), 6(b), and 7 present a comprehensive comparative analysis of time transfer stability between HCF and SMF over a 68-km link. Acquired using identical hardware under a continuous 24-hour diurnal cycle, the data unequivocally demonstrate the superior long-term robustness of the HCF medium.

First, Fig. 7(a) characterizes the residual time deviation. Throughout the test duration, the HCF link maintains a tightly constrained deviation of < 100 ps, with a residual peak-to-peak fluctuation of 50 ps. In contrast, while the SMF link exhibits a comparable short-term fluctuation amplitude (50 ps), its absolute time deviation settles at a significantly higher baseline of ~ 400 ps. This distinction reveals that under identical environmental conditions, the systematic timing offset in HCF is reduced by a factor of four, reflecting its inherent immunity to external perturbations.

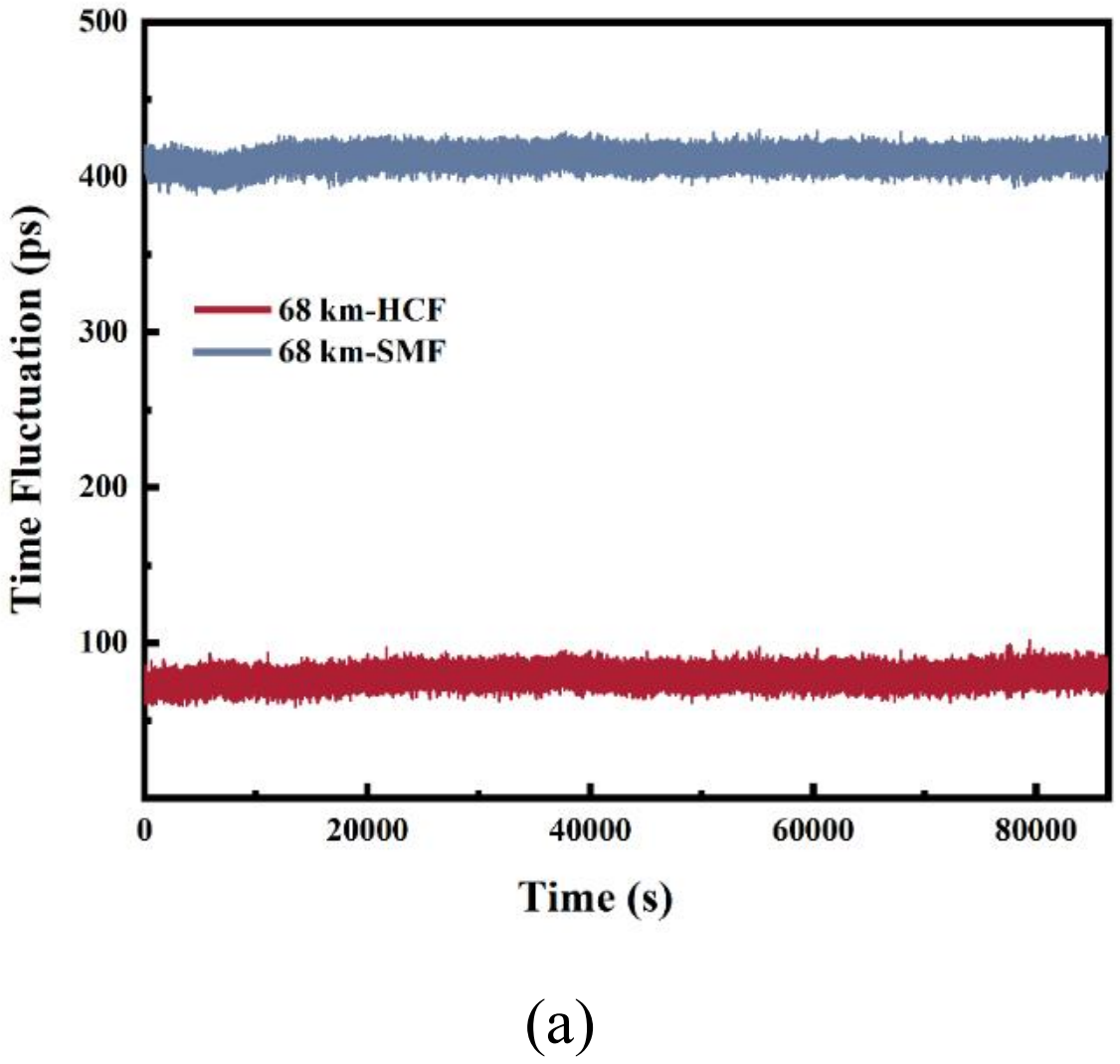

(a)

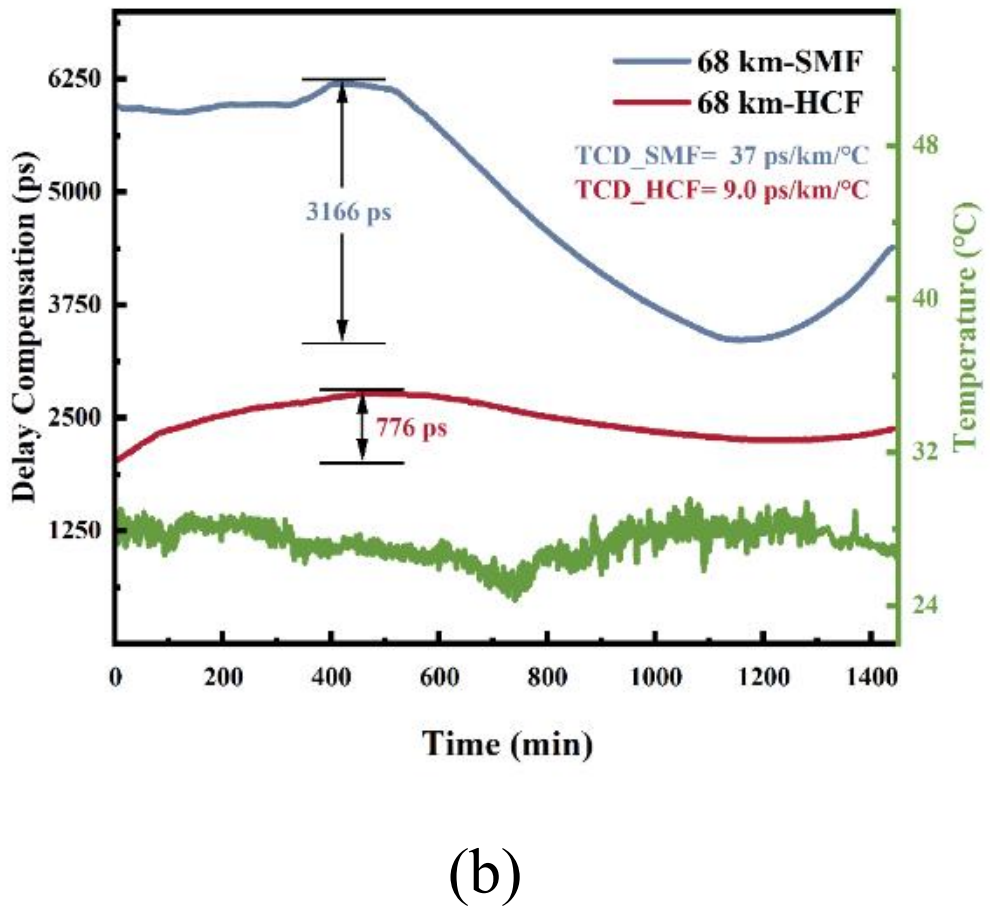

(b)

**Fig. 7.** Comparison of 24-hour time transfer stability over 68-km links. (a) Time delay fluctuations for HCF and SMF. (b) Real-time delay compensation for HCF and SMF links.

Crucially, Fig. 7(b) substantiates the thermal insensitivity of HCF by plotting the real-time link delay drift. Driven by diurnal temperature cycling, the SMF link undergoes a delay drift of 3166 ps. Given the ambient fluctuation of ~5 ℃, this magnitude reflects an attenuated effective thermal load on the fiber spools attributed to their significant thermal inertia. In contrast, the HCF link, subjected to the identical thermal environment, exhibits a drift of only 776 ps. This fourfold suppression aligns precisely with the ratio of the intrinsic thermal coefficients of delay (TCD) of the two fibers (~37 ps/km/℃ for SMF vs. ~9 ps/km/℃ for HCF). Such consistency validates HCF as a highly stable medium for precision time transfer, significantly mitigating thermally induced latency variations.

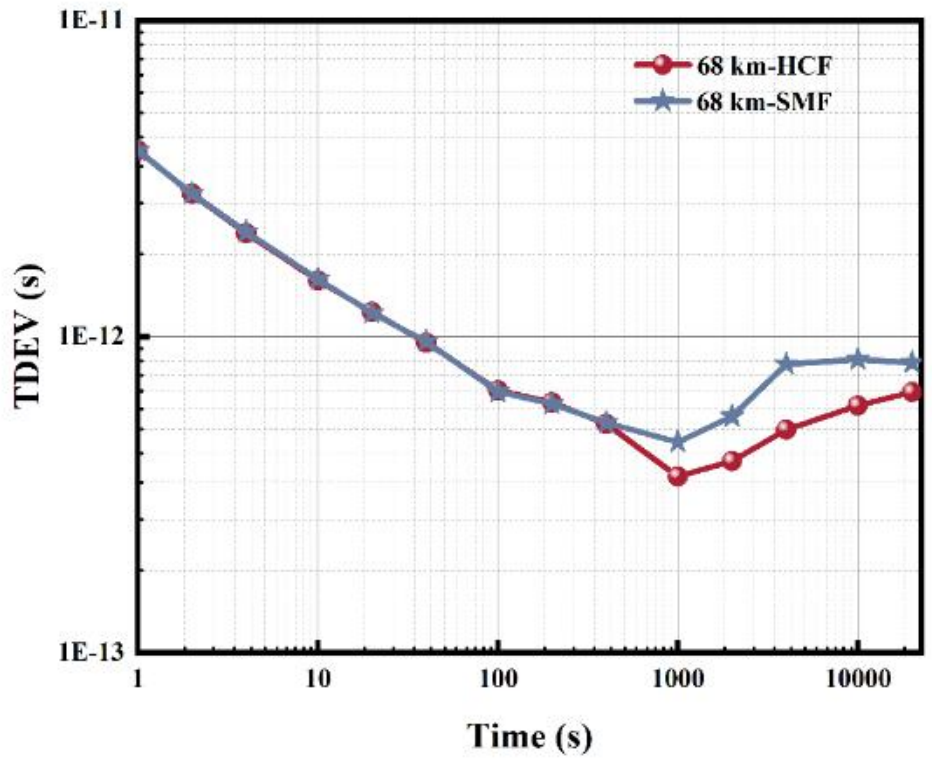

**Fig. 8.** Analysis of Time Deviation (TDEV) stability limits.

The Time Deviation (TDEV) analysis in Fig. 8 further elucidates the distinct performance characteristics of the two links across varying timescales. In the short-term integration regime ( $1\,s < \tau < \sim 200\,s$ ), the TDEV profiles exhibit near-perfect overlap. This coincidence confirms that stability in this domain is dictated by the common-mode noise floor of the measurement instrumentation—including lasers, photodetectors, and time interval counters—rather than the fiber medium itself. Conversely, in the medium-to-long-term regime ( $\tau > 200\,s$ ), where environmental coupling mechanisms become dominant, the performance trajectories diverge significantly. The HCF link (black curve) demonstrates continuous stability enhancement, achieving a floor of 0.2 ps at $\tau$=1000 s before a gentle ascent induced by residual long-term drift. In contrast, the SMF link (red curve) reaches a minimum of only 0.4 ps at 1000 s. More critically, beyond this optimum, the SMF curve exhibits a steep upward inflection, indicating a susceptibility to long-term environmental drift that is substantially more severe than that of HCF.

This 24-hour comparative study provides compelling empirical evidence of a fundamental performance shift: while short-term precision remains instrument-limited, the medium-to-long-term stability—the definitive bottleneck for long-haul time transfer—is governed by the transmission medium. Here, HCF leverages its intrinsic low dispersion and environmental decoupling to deliver superior absolute timing stability. Consequently, it stands as the premier transmission medium for enabling the next generation of ultra-high-precision, ultra-stable long-haul time-frequency distribution networks.

## V. CONCLUSION

In summary, we have experimentally validated the superior robustness of hollow-core fiber for long-haul high-precision time transfer. By systematically characterizing links up to 91 km, we confirm that HCF suppresses dispersion-induced timing errors and Kerr-induced phase noise by orders of magnitude compared to SMF. The achieved TDEV of 0.2 ps at 1000 s over 68 km, combined with a fourfold reduction in thermal sensitivity, highlights the potential of HCF to simplify network architectures by removing the need for complex active compensation schemes. Leveraging the low thermal sensitivity of Hollow Core Fiber (HCF), direct time signal distribution without complex feedback control holds promise for meeting diverse ordinary clock synchronization requirements. Furthermore, the reduced environmental enhances the practicality of optical fiber timing. These results pave the way for simplified, ultra-stable optical clock distribution networks essential for next-generation quantum networking and deep-space navigation.

## ACKNOWLEDGMENT

This research was supported by the National Major Science and Technology Infrastructure Project of China "High Precision Ground-based Time Service System" (2017-000052-73-01-002401); The special research assistant funding project, CAS (No. 110100T0LB); Shaanxi Provincial Postdoctoral Science Foundation (E461RC1101);National Natural Science Foundation of China (Grant Nos. 12303077 and 62120106010); Innovation Program for Quantum Science and Technology (Grant Nos. 2021ZD0300900). All authors have reviewed and approved the final version of the manuscript.

(Corresponding author: Tao Liu, Ruifang Dong) Bo Liu and Xinxing Guo contributed to the work equally and should be regarded as co-first authors.